\documentclass[twocolumn,aps,prl,showpacs,superscriptaddress,preprintnumbers,floatfix,nofootinbib,longbibliography]{revtex4-1}
\usepackage{url}
\usepackage{graphicx}
\usepackage{dcolumn}
\usepackage{bm}
\usepackage[usenames]{color}
\usepackage{amssymb}
\usepackage{amsmath}
\usepackage{multirow}
\usepackage{harpoon}
\usepackage{MnSymbol}
\usepackage{appendix}

\usepackage{physics}
\usepackage{cancel}
\usepackage{bm}
\usepackage{capt-of}
\usepackage{xcolor}
\usepackage{booktabs}
\usepackage{siunitx}
\usepackage{placeins}
\usepackage[colorlinks,linkcolor=blue!79,citecolor=blue!79,filecolor=blue!79,urlcolor=blue!79]{hyperref}

\begin{document}

\title{Scaling approach to rigid and soft nuclear deformation through flow fluctuations in high-energy nuclear collisions}

\newcommand{\moe}{Key Laboratory of Nuclear Physics and Ion-beam Application (MOE), and Institute of Modern Physics, Fudan
University, Shanghai 200433, China}
\newcommand{\fudan}{Shanghai Research Center for Theoretical Nuclear Physics, NSFC and Fudan University, Shanghai 200438, China}
\newcommand{\fudanP}{Physics Department and Center for Particle Physics and Field Theory, Fudan University, Shanghai 200438, China}
\newcommand{\sbu}{Department of Chemistry, Stony Brook University, Stony Brook, NY 11794, USA}
\newcommand{\bnl}{Physics Department, Brookhaven National Laboratory, Upton, NY 11976, USA}

\author{\small Lumeng Liu}\affiliation{\fudanP}
\author{\small Chunjian Zhang}\email{chunjianzhang@fudan.edu.cn}\affiliation{\moe}\affiliation{\fudan}
\author{\small Jinhui Chen}\email{chenjinhui@fudan.edu.cn}\affiliation{\moe}\affiliation{\fudan}
\author{\small Jiangyong Jia}\email{jiangyong.jia@stonybrook.edu}\affiliation{\sbu}\affiliation{\bnl}
\author{\small Xu-Guang Huang}\email{huangxuguang@fudan.edu.cn}\affiliation{\fudanP}\affiliation{\moe}\affiliation{\fudan}
\author{\small Yu-Gang Ma}\email{mayugang@fudan.edu.cn}\affiliation{\moe}\affiliation{\fudan}

\begin{abstract}
The nature of octupole deformation, whether static or vibrational, remains an open question in nuclear physics. Here, we propose a scaling approach to probe this ambiguity by triangular flow fluctuations using multi-particle cumulants, $c_{3,\varepsilon}\{4\}$, in relativistic $^{238}$U+$^{238}$U collisions. We demonstrate that both $|c_{3,\varepsilon}\{4\}|$ and the ratio $|c_{3,\varepsilon}\{4\}/c^2_{3,\varepsilon}\{2\}|$ scale linearly with the fourth-order moment of octupole deformation, $\langle \beta^4_{3,\mathrm{U}} \rangle$. Combined with the known linear relation of $c_{3,\varepsilon}\{2\}$ to $\langle \beta^2_{3,\mathrm{U}} \rangle$, this new relation provides a direct extraction of both the mean and variance of the octupole deformation fluctuations, finally discriminating between static and dynamic origins. This work establishes a new tool to probe the static and dynamic collective modes in high-energy nuclear collisions, advancing a significant step toward refining the initial conditions of quark-gluon plasma.
\end{abstract}
\maketitle

\textit{Introduction.---}
The quest to understand the fundamental shapes of atomic nuclei and their collective dynamics is a central theme in nuclear physics. Octupole deformation, characterized by a ``pear-like" shape reflecting asymmetric mass distribution, has garnered sustained interest for decades~\cite{Verney:2025efj}. Its presence is conceivably linked to fundamental phenomena such as the emergence of static electric dipole moments (EDM) and the breaking of parity symmetry in the nuclear landscape~\cite{Ahmad:1993nx,RevModPhys.91.015001}. The heavy, strongly quadrupole-deformed nuclei stand as a prime candidate in the search for stable octupole collectivity~\cite{Gaffney2013,Butler:2016rmu}. Compelling evidence for non-zero octupole correlations in $^{238}$U has been accumulated from low-energy experimental signatures~\cite{Spear:1989zz,MCGOWAN1994569,KIBEDI:2002wxc}. A core challenge lies in distinguishing the intrinsic nature of these correlations: do they manifest as a static, rigid deformation of the nuclear ground state, or do they arise from soft vibrational modes that dynamically fluctuate around a mean spherical or deformed shape? Conventional low-energy probes, such as the study of excited state parities and transition rates, face inherent challenges in distinguishing these scenarios, as both can produce qualitatively similar spectra~\cite{Agbemava:2017vug}.

In recent years, a novel approach has emerged: using relativistic heavy-ion collisions as a microscope for nuclear structure by taking advantage of the strong responses of the hydrodynamics collective flow to the shape and size of the initial-state quark-gluon plasma (QGP)~\cite{Jia:2022ozr,Jia:2021qyu,Ma:2022dbh,Jia:2022qgl,ChunJian:2024vdk,Giacalone:2025vxa,Luzum:2023gwy,Shou:2014eya,Duguet:2025hwi,Jia:2025wey,Ke:2025tyv}. This approach has been applied to study the quadrupole deformation $\beta_2$~\cite{Schenke:2020mbo,Giacalone:2021udy,STAR:2024wgy,Giacalone:2024bud,STAR:2025vbp,Fortier:2023xxy,Fortier:2024yxs}, octupole deformation $\beta_3$~\cite{Carzon:2020xwp,Zhang:2021kxj,Samanta:2023qem,STAR:2025vbp,Zhang:2025hvi,Wang:2024ulq,LHCb:2025ixz}, hexadecapole deformation $\beta_4$~\cite{Ryssens:2023fkv,Xu:2024bdh,Wang:2024vjf,Bjorn}, the triaxial deformation $\gamma$~\cite{Bally:2021qys,ATLAS:2022dov,ALICE:2021gxt,Lu:2023fqd,Zhao:2024lpc}, neutron skin~\cite{Xu:2021vpn,Jia:2022qgl,Jia:2021oyt,Giacalone:2023cet,YuGangMa}, and the possible alpha cluster structure~\cite{STAR:2025ivi,ATLAS:2025nnt,ALICE:2025luc,LHCb:2025ixz} in the ground-state nuclei. The event-by-event initial geometric anisotropy of the colliding nuclei is converted via the strong interaction into momentum anisotropy of the final-state particles at RHIC and the LHC~\cite{Chen:2024aom,Shou:2024uga}, quantified by harmonic flow coefficients, characterized by a Fourier expansion $\mathrm{d} N / \mathrm{d} \phi \propto 1+2 \sum_{n=1}^{\infty} v_n \cos n\left(\phi-\Phi_n\right)$, where $v_n$ and $\Phi_n$ represent the magnitude and event-plane angle of the $n^{\text {th}}$-order harmonic flow~\cite{Shuryak:1999by}. Notably, the anisotropic flow $\langle v_n^2 \rangle$, has been established as a sensitive probe of nuclear deformation. This proportionality holds since $\langle v_n^2 \rangle$ scales linearly with the mean squared eccentricity $\langle \varepsilon_n^2 \rangle$, which in turn reflects the second-order moment of the deformation parameter, $\langle \beta_n^2 \rangle$~\cite{Jia:2021tzt,Jia:2021qyu}.

The recent analysis of $^{238}$U+$^{238}$U collisions at RHIC energies provided compelling evidence for a non-zero average octupole deformation through measurements of two- and three-particle correlations, $\langle v_3^2\rangle$ and $\langle v_3^2\delta p_{\rm T}\rangle$~\cite{STAR:2025vbp}. These observables primarily constrain the mean square deformation $\langle \beta_3^2 \rangle$, which is insensitive to the critical distinction between a static deformation (narrow distribution of $\beta_3$) and a dynamic vibration (wide distribution of $\beta_3$). As anticipated, event-by-event flow fluctuations, quantified by the standard deviation $\sigma_{v_n}$ of $v_n$ distribution in central collisions, are strongly sensitive to the fluctuations of the spatial orientation of the colliding nuclei~\cite{Bhalerao:2014xra,Giacalone:2018apa,Bhalerao:2018anl,Jia:2022qgl}. In particular, this phenomenon of nuclear shape fluctuations has attracted considerable attention in $^{208}$Pb nucleus for solving the high-energy ultra-central $v_2$-to-$v_3$ puzzle~\cite{Zakharov:2020irp,Zakharov:2021lux,Xu:2025cgx} and the low-energy heavy-ion fusion reactions~\cite{Hagino:2025vxe}.  

In this work, we present a scaling approach to probe nuclear shape fluctuations in high-energy nuclear collisions. An analytical framework that connects higher-order moments of the nuclear deformation distribution to multi-particle correlation observables is applied to the ground-state heavy $^{238}$U nucleus. Specifically, we demonstrate that the triangular flow fourth-order cumulant $|c_{3,\varepsilon}\{4\}|$ and the double ratio 
$|R_{c_{3,\varepsilon}\{4\}/c_{3,\varepsilon}^2\{2\}}|$, defined as the ratio of $c_{3,\varepsilon}\{4\}/c_{3,\varepsilon}^2\{2\}$ between two collision systems, exhibit an approximately linear dependence on the fourth-order moment $\langle \beta_3^4 \rangle$. Combined with the established linear relation of $c_{3,\varepsilon}\{2\}$ to $\langle \beta_3^2 \rangle$, this new relation allows us to simultaneously extract both the mean and the variance of the octupole deformation distribution. Applying this method to existing high-energy $^{238}$U collision data will provide the first direct evidence to distinguish the nature of octupole collective mode, thereby answering a fundamental question in nuclear structure through a unique high-energy collision experiment.

\textit{Linking shape fluctuations to four-particle correlations.---} 
We introduce a novel experimental observable to discriminate between soft vibrational and rigid octupole deformation in nuclei~\cite{Jia:2021tzt,Jia:2021qyu}. Our approach derives from analytical relations connecting the second- and fourth-order cumulants of the anisotropic eccentricity, $\varepsilon_n\{2\}$ and $\varepsilon_n\{4\}$, to the second- and fourth-order moments of the deformation parameter $\beta_n$, i.e., $\langle \beta_n^2 \rangle$ and $\langle \beta_n^4 \rangle$. These moments directly constrain the mean deformation, $\bar{\beta}_n$, and its variance, $\sigma_{\beta_n}^2$, under the assumption of Gaussian fluctuations.

We start with the eccentricity vector $\vec{\varepsilon}_{n} = \varepsilon_{n} e^{i n \Phi_{n}}$, an estimator for the anisotropic flow $v_n e^{i n \Psi_{n}}$, which is computed from event-by-event transverse entropy density $s(\vec{r}_\perp)$ using
\begin{align}\label{eq:eccn}
\vec{\varepsilon}_{n}
&= -\frac{\displaystyle \int r_\perp^{n} \, e^{i n \phi} \, s(\vec{r}_\perp) \, d^{2}\vec{r}_\perp}
{\displaystyle \int r_\perp^{n} \, s(\vec{r}_\perp) \, d^{2}\vec{r}_\perp},
\end{align}
Here, $r_\perp^{n}$ denotes the transverse radius raised to the power $n$. In ultra-central collisions of nuclei with small intrinsic deformations $\beta_n$, the eccentricity vector can be expressed to leading order in $\beta_n$ as~\cite{Jia:2021tzt,Jia:2021qyu,Dimri:2023wup}
\begin{align} \label{eq:epsilonn}
\vec{\varepsilon}_{n} = \vec{\varepsilon}_{n,0} + \vec{p}_{n}(\Omega) \beta_n + \mathcal{O}(\beta_n^2),
\end{align}
where $\vec{\varepsilon}_{n,0}$ represents the eccentricity contribution originating from spherical nuclei, and the phase space factor $\vec{p}_{n}(\Omega) = - A_{n} \sum_m \alpha_{n,m} D^{n}_{n,m}(\Omega)$ depends on the Euler rotation angles $\Omega$. Here, $D_{n,m}^{l}(\Omega)$ denotes the Wigner matrix, $\alpha_{n,m}$ represents the deformation components for different $m$, and $A_n$ is numerically $A_{n} \equiv \frac{(n+3) \, \Gamma\left(1 + \frac{1}{2} + \frac{n}{2}\right)}{\pi \, \Gamma\left(1 + \frac{n}{2}\right)} 
\sqrt{\frac{(2n)!!}{(2n+1)!!}}$. The first term on the right-hand side of Eq.~\eqref{eq:epsilonn}, $\vec{\varepsilon}_{n,0}$, has distinct origins for different harmonic orders: for elliptic flow ($n=2$), it arises from both average collision geometry and nucleon fluctuations, while for triangular flow ($n=3$), it originates solely from nucleon fluctuations due to the symmetry under target-projectile exchange, which ensures $\left\langle \vec{\varepsilon}_{3} \right\rangle = 0$. The second term, proportional to $\beta_n$,  emerges from the nuclear deformation and can be evaluated by modeling the nucleus as a uniform sharp sphere with surface radius deformation 
\(
R(\theta,\phi) = R_0 \left[ 1 + \sum_{l,m} \beta_l \alpha_{l,m} Y_{l,m}(\theta,\phi) \right].
\)

We usually measure anisotropic flow using two-particle correlation method, and this second-order cumulant of the flow vector decomposes into a squared mean and variance as
$\left\langle \varepsilon_n^2 \right\rangle = \left\langle \varepsilon_n \right\rangle^2 + \sigma^2_{\varepsilon_n}$. Assuming the event-by-event fluctuations of $\vec{\varepsilon}_{n,0}$, $\beta_n$, and the rotation matrix are uncorrelated, the event-averaged two-particle correlation simplifies to
\begin{align}\label{eq:cn2} 
c_{n,\varepsilon}\{2\} = \left\langle \varepsilon_n^2 \right\rangle \approx \left\langle \varepsilon_{n,0}^2 \right\rangle + \left\langle p_n p_n^* \right\rangle \left\langle \beta_n^2 \right\rangle,
\end{align}
The fourth-order cumulant, $c_{n,\varepsilon}\{4\}$, is widely employed to suppress the effects of flow fluctuations~\cite{Bhalerao:2006tp,Voloshin:2007pc}, thereby enabling a more accurate extraction of the reaction-plane flow, $\langle \varepsilon_n \rangle$, from experimental measurements. In the presence of nuclear deformation, additional contributions from both shape and orientation fluctuations arise. The influence of initial-state deformation on $c_{n,\varepsilon}\{4\}$ can be expressed as
\begin{align} \label{eq:cn4} 
c_{n,\varepsilon}\{4\} &= \left\langle \varepsilon_n^4 \right\rangle - 2 \left\langle \varepsilon_n^2 \right\rangle^2 \nonumber\\
&\approx \left\langle \varepsilon_{n,0}^4 \right\rangle - 2 \left\langle \varepsilon_{n,0}^2 \right\rangle^2 
+ \left\langle p_n^2 p_n^{2*} \right\rangle \left\langle \beta_n^4 \right\rangle 
- 2 \left\langle p_n p_n^* \right\rangle^2 \left\langle \beta_n^2 \right\rangle^2.
\end{align}
Specifically, for triangular flow ($n=3$) with octupole deformation ($\beta_3$), the contribution from nucleon fluctuations, $\left\langle \varepsilon_{n,0}^4 \right\rangle - 2 \left\langle \varepsilon_{n,0}^2 \right\rangle^2$, is subdominant compared to the deformation-induced term $
\left\langle p_3^2 p_3^{2*} \right\rangle \left\langle \beta_3^4 \right\rangle - 2 \left\langle p_3 p_3^* \right\rangle^2 \left\langle \beta_3^2 \right\rangle^2$, which is sensitive to both the magnitude and the event-by-event fluctuations of $\beta_3$. This term scales approximately linearly with the fourth-order moment $\left\langle \beta_3^4 \right\rangle$ when the second-order moment $\left\langle \beta_3^2 \right\rangle = \bar{\beta}_3^2 + \sigma_{\beta_3}^2$ is fixed, thereby providing direct access to the variance of the octupole deformation.

In our previous work~\cite{STAR:2025vbp,Zhang:2025hvi}, compelling evidence for octupole deformation in $^{238}$U was established through comparative measurements of the two-particle triangular flow $\langle v_3^2 \rangle$ in relativistic $^{238}$U+$^{238}$U and $^{197}$Au+$^{197}$Au collisions. The results idicate a small but nonzero octupole deformation, with $\beta_{\rm 3,U} \sim 0.08-0.1$. However, the origin of this deformation—specifically, whether it originates from soft vibrational excitations or a static rigid deformation—remains an open question, as the two-particle cumulant $\left\langle v_3^2 \right\rangle$ probes only the second-order moment $\left\langle \beta_3^2 \right\rangle = \bar{\beta}_3^2 + \sigma_{\beta_3}^2$, as shown in Eq.~\eqref{eq:cn2}. Hence, the constrait available from this observable is therefore given to $\left\langle \beta_\mathrm{3,U}^2 \right\rangle = \bar{\beta}_\mathrm{3,U}^2 + \sigma_{\beta_\mathrm{3,U}}^2 \approx 0.01$. We now take a significant
step forward by exploring the fourth-order cumulant $c_{3,\varepsilon}\{4\}$, defined in Eq.~\eqref{eq:cn4}, together with $c_{3,\varepsilon}\{2\}$, to separately extract the mean deformation $\bar{\beta}_\mathrm{3,U}$ and its fluctuation $\sigma_{\beta_\mathrm{3,U}}$. This approach enables a clear discrimination between vibrational and static octupole collectivity.

For simplicity, in this study, we focus on the dominant axial component $\alpha_{3,m}$ ($m=0$) of octupole deformation and assume that the deformation parameters undergo Gaussian fluctuations, characterized by a mean $\bar{\beta}_3$ and standard deviation $\sigma_{\beta_3}$. In the body frame of an axially symmetric nucleus, the octupole deformation is conventionally parameterized by the $m=0$ term alone, and the deformation parameter $\beta_3$ entering the Woods-Saxon potential is defined accordingly. Terms with $m \neq 0$ correspond to non-axially-symmetric octupole modes, which are subdominant for $^{238}$U and are neglected here~\cite{bohr}. Under these assumptions, $\bar{\beta}_3$ and $\sigma_{\beta_3}$ can be expressed as
\begin{align} 
\begin{split}
\bar{\beta}_3 &= \sqrt[4]{\frac{3 \left\langle \beta_3^2 \right\rangle^2 - \left\langle \beta_3^4 \right\rangle}{2}},\\
\sigma_{\beta_3} &= \sqrt{ \left\langle \beta_3^2 \right\rangle - \sqrt{ \frac{3 \left\langle \beta_3^2 \right\rangle^2 - \left\langle \beta_3^4 \right\rangle}{2} } }. \label{eq:beta3var}
\end{split}
\end{align}
Fluctuations beyond the Gaussian approximation—specifically, the skewness and kurtosis of $\beta_{n}$—will be briefly addressed in the concluding section of the paper.

\textit{Model simulations.---} To evaluate the sensitivity of these flow cumulants to octupole deformation and its fluctuations, we implement these derived relations into the \texttt{T\raisebox{-.5ex}{R}ENTo} initial condition model~\cite{Moreland:2014oya}. By systematically varying the mean deformation $\bar{\beta}_3$ and its flucutation $\sigma_{\beta_3}$ under the constraint of a fixed second-order moment $\langle \beta_3^2 \rangle$, we demonstrate that the fourth-order cumulant offers essential additional constraints for disentangling these two deformation parameters.

In this \texttt{T\raisebox{-.5ex}{R}ENTo} model, the entropy density profile is modeled by $s(\vec{r}_\perp)\propto\sqrt{T_A(\vec{r}_\perp+\vec{b}/2) T_B(\vec{r}_\perp-\vec{b}/2)}$ with $T_A$ and $T_B$ are the thickness functions of the colliding nuclei. In the present calculation, we set the reduced thickness parameter $p=0$, the nucleon width $w=0.5$ fm, the number of constituents in each nucleon $n_c=1$, and the minimum nucleon-nucleon distance $d_\text{min}=0.9$ fm.
The nucleon distribution is described by a deformed Woods-Saxon form:
\begin{equation}\label{WS}
\rho(r,\theta,\phi) = \frac{\rho_0}{1 + \exp\left(\frac{r - R(\theta,\phi)}{a}\right)},
\end{equation}

\begin{table}[htbp]
    \centering
    \caption{Octupole deformation parameters used in the simulations of $^{238}$U +$^{238}$U collisions. ``$\rm U_{case0}$" and ``$\rm U_{case6}$" represent solely the octupole deformation variance $\sigma_{\beta_3}=0.01$ (soft) and the mean deformation $\bar{\beta}_3=0.01$ (rigid), respectively.}
    \label{tab:beta3}
    \renewcommand{\arraystretch}{1.3}
    \setlength{\tabcolsep}{4mm}
    \begin{tabular}{ccccc}
        \hline\hline
        Case & $\bar{\beta}_3$ & $\sigma_{\beta_3}$ & $\langle \beta_3^2 \rangle$ & $\langle \beta_3^4 \rangle$ \\
        \hline
        $\text{U}_{\rm case 0}$ & 0.00 & 0.100 & 0.01 & 0.000300 \\
        $\text{U}_{\rm case 1}$ & 0.05 & 0.087 & 0.01 & 0.000288 \\
        $\text{U}_{\rm case 2}$ & 0.06 & 0.080 & 0.01 & 0.000274 \\
        $\text{U}_{\rm case 3}$ & 0.07 & 0.071 & 0.01 & 0.000252 \\
        $\text{U}_{\rm case 4}$ & 0.08 & 0.060 & 0.01 & 0.000218 \\
        $\text{U}_{\rm case 5}$ & 0.09 & 0.044 & 0.01 & 0.000169 \\
        $\text{U}_{\rm case 6}$ & 0.10 & 0.000 & 0.01 & 0.000100 \\
        \hline\hline
    \end{tabular}
\end{table}

The primary goal of this work is to establish the universal scaling approach and also facilitate its validation through experimental measurements. Thus, we focus on determining the mean octupole deformation $\bar{\beta}_3$ and its variance $\sigma_{\beta_3}$ for $^{238}$U, while fixing other deformation parameters to the values consistent with prior studies. For $^{238}$U, we use a nuclear radius $R_{\rm 0,U}=6.81$ fm, surface diffuseness $a_{\rm U}=0.55$ fm, and quadrupole and hexadecapole deformations $\beta_{\rm 2, U}=0.28$ and $\beta_{\rm 4, U}=0.09$. For $^{197}$Au, the parameters are $R_{\rm 0,Au}=6.62$ fm, $a_{\rm AU}=0.52$ fm, $\beta_{\rm 2, Au}=0.14$, $\gamma_{\rm Au}=45^{\circ}$, as adopted in Refs.~\cite{STAR:2024wgy,STAR:2025vbp}. For each value of \(\langle\beta_{3,\mathrm{U}}^4\rangle\) in the deformation scan, we generate 2 $\times$ $10^8$ minimum-bias \texttt{T\raisebox{-.5ex}{R}ENTo} events for $^{238}$U+$^{238}$U collisions. The $^{197}$Au+$^{197}$Au reference system is generated with the same statistics. Centrality is determined according to the total initial entropy. Thus, for each deformation point, the 0--2\% and 0--5\% centrality intervals contain approximately \(4\times10^6\) and \(10^7\) events, respectively. Such high statistics is required to obtain stable estimates of the small four-particle eccentricity cumulants in ultra-central collisions.

\begin{figure}[htbp]
\centering
\includegraphics[width=0.75\linewidth]{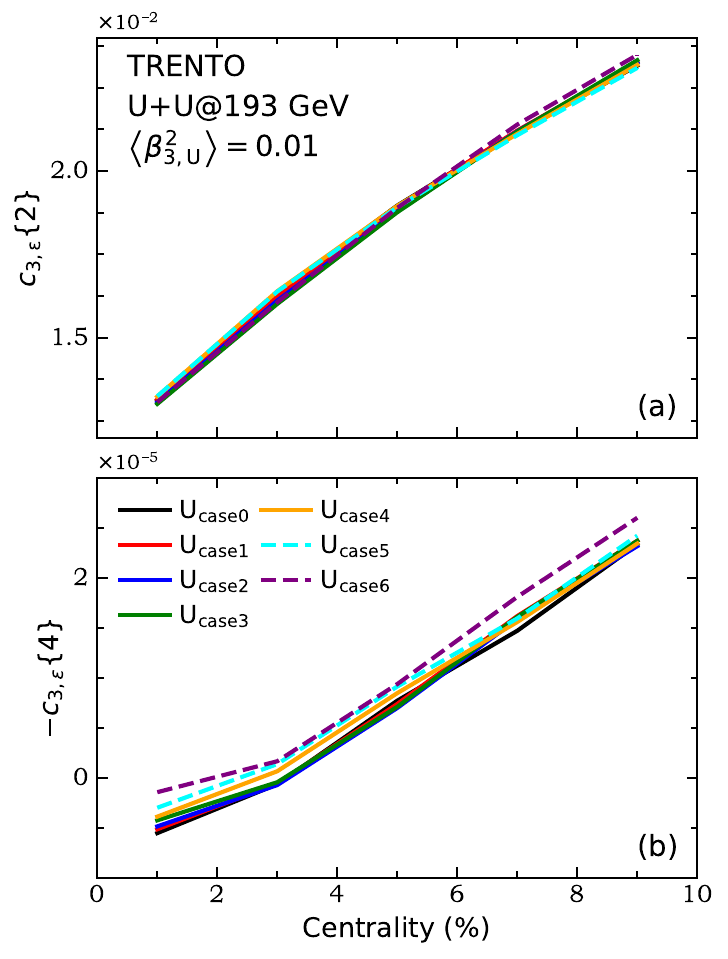}
\vspace{-0.2cm}
\caption{Centrality dependence of the two-particle cumulant $c_{3,\varepsilon}\{2\}$ (a) and four-particle cumulant $c_{3,\varepsilon}\{4\}$ (b) from \texttt{T\raisebox{-.5ex}{R}ENTo} model are presented for $^{238}$U+$^{238}$U collisions at 193 GeV. The results are obtained with a fixed value of $\langle \beta_\mathrm{3,U}^2\rangle$ but varying mean octupole deformations $\bar{\beta}_{\rm 3,U}$ and variances $\sigma_{\beta_{\rm 3,U}}^2$, as specified in Tab.~\ref{tab:beta3}.}
\label{fig:c32}
\end{figure}

\textit{Results and discussions.---} 
\begin{figure}[b]
\centering
\includegraphics[width=0.95\linewidth]{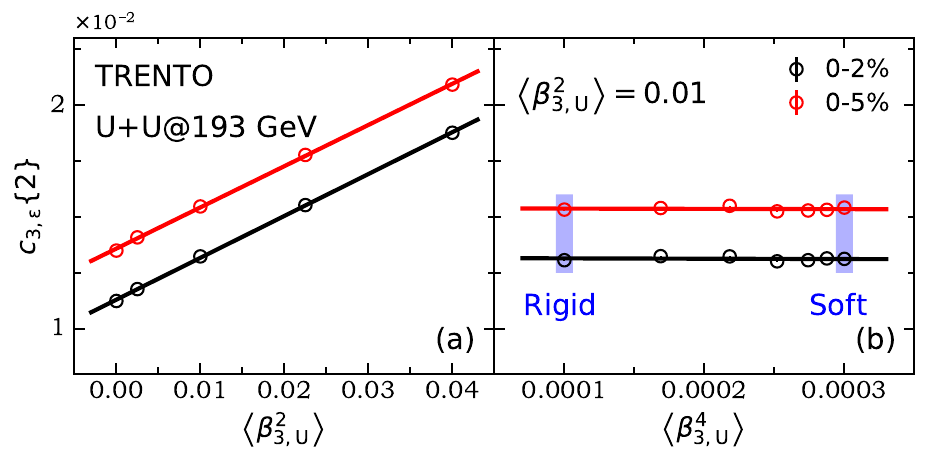}
\vspace{-0.2cm}
\caption{Dependence of the two-particle cumulant $|c_{3,\varepsilon}\{2\}|$ on the second-order moment $\langle \beta_\mathrm{3,U}^2 \rangle$ (panel a) and fourth-order moment $\langle \beta_\mathrm{3,U}^4 \rangle$ (panel b) from \texttt{T\raisebox{-.5ex}{R}ENTo} model in $^{238}$U+$^{238}$U collisions with two centrality intervals, 0--2\% (black symbols) and 0--5\% (red symbols). Linear fits are shown as solid lines. The light blue bands depict exclusively the rigid and soft deformation cases.}
\label{fig:c32beta}
\end{figure}
Figure~\ref{fig:c32}(a) and (b) show the centrality dependence of the two-particle cumulant $c_{3,\varepsilon}\{2\}$ and the four-particle cumulant $c_{3,\varepsilon}\{4\}$, respectively, in $^{238}$U+$^{238}$U collisions at 193 GeV. The results are obtained for various combinations of $(\bar{\beta}_\mathrm{3,U}, \sigma_{\beta_\mathrm{3,U}})$ that yield different values of $\langle \beta_\mathrm{3,U}^4 \rangle$, while keeping the second-order moment $\langle \beta_\mathrm{3,U}^2 \rangle$ fixed at 0.01. As expected, the two-particle correlation for the initial-state triangular eccentricity, $\langle \varepsilon_3^2 \rangle$ (or equivalently, the final-state two-particle triangular flow, $\langle v_3^2 \rangle$), depends only on the second-order moment of the octupole deformation and exhibits no sensitivity to the fourth-order moment, consistent with Eq.\eqref{eq:cn2}. In contrast, $c_{3,\varepsilon}\{4\}$ exhibits a clear hierarchical dependence on $\langle \beta_\mathrm{3,U}^4 \rangle$, as shown in Fig.\ref{fig:c32}(b).

Figure~\ref{fig:c32beta}(a) demonstrates that $c_{3,\varepsilon}\{2\}$ is indeed linear in $\langle \beta_{\rm 3,U}^2 \rangle$ across two different centrality class, consistent with our earlier findings~\cite{Giacalone:2021udy,Zhang:2021kxj,Jia:2021oyt}. However, we find that $c_{3,\varepsilon}\{2\}$ exhibits no sensitivity to $\langle \beta_{\rm 3,U}^4 \rangle$ when $\langle \beta_\mathrm{3,U}^2 \rangle$ is fixed, as expected in Fig.~\ref{fig:c32beta}(b).

\begin{figure}[t]
\centering
\includegraphics[width=0.75\linewidth]{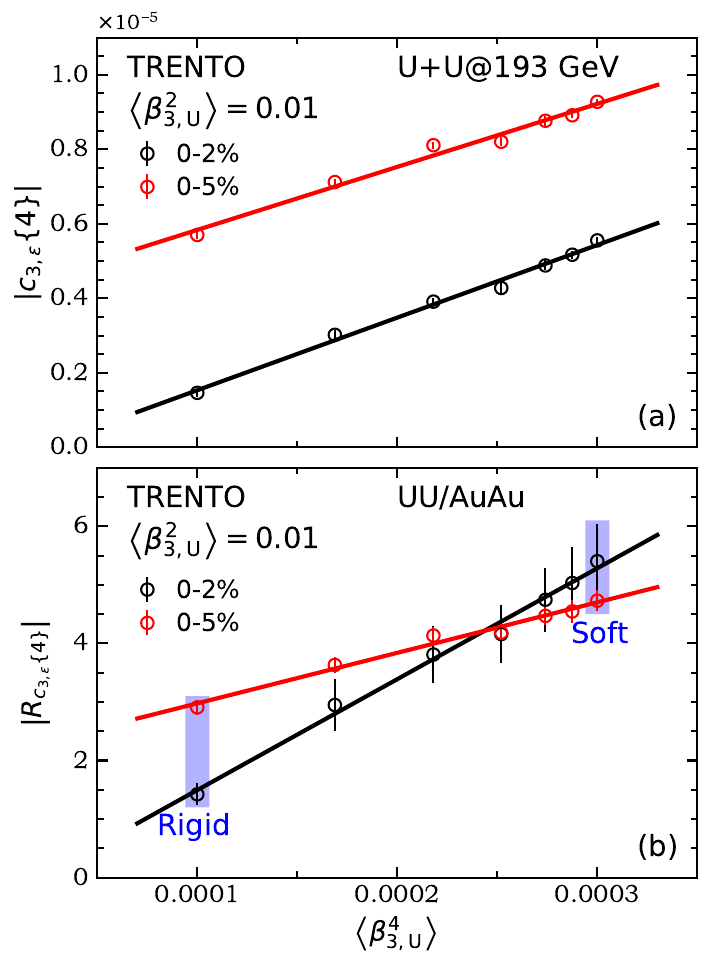}
\vspace{-0.2cm}
\caption{(a) Dependence of the four-particle cumulant $|c_{3,\varepsilon}\{4\}|$ on the fourth-order moment $\langle \beta_\mathrm{3,U}^4 \rangle$ from \texttt{T\raisebox{-.5ex}{R}ENTo} model in two different centrality intervals, 0--2\% (black symbols) and 0--5\% (red symbols). (b) The ratio between $^{238}$U+$^{238}$U and $^{197}$Au+$^{197}$Au collisions, $|R_{c_{3,\varepsilon}\{4\}}|$, as a function of $\langle \beta_\mathrm{3,U}^4 \rangle$. Linear fits are shown as solid lines. The light blue bands depict exclusively the rigid and soft deformation cases.}
\label{fig:c34ratio}
\end{figure}

To isolate the effects of $\beta_3$ fluctuations, Fig.~\ref{fig:c34ratio}(a) presents the dependence of the four-particle eccentricity cumulant $|c_{3,\varepsilon}\{4\}|$ on the fourth-order moment $\langle \beta_\mathrm{3,U}^4 \rangle$ for two most central collisions, 0--2\% and 0--5\%. The results show an approximately linear increase of 
\(|c_{3,\varepsilon}\{4\}|\) with 
\(\langle \beta_{\mathrm{3,U}}^4 \rangle\) at fixed 
\(\langle \beta_{\mathrm{3,U}}^2 \rangle\), consistent with Eq.~\eqref{eq:cn4}. Under the linear-response approximation, \(v_3=\kappa_3\varepsilon_3\), this trend is expected to be reflected in the final-state four-particle flow cumulant 
\(|c_{3,v}\{4\}|=|v_3^4\{4\}|\), up to an overall response factor \(\kappa_3^4\).

To further reduce final-state effects, we calculate the ratio of an observable $\mathcal{O}$ between isobar-like collisions systems $^{238}$U+$^{238}$U at 193 GeV and $^{197}$Au+$^{197}$Au at 200 GeV at a given centrality~\cite{Giacalone:2021uhj},
\begin{align} \label{eq:R}
R_{\mathcal{O}} = \frac{\mathcal{O}_{\rm{U+U}}}{\mathcal{O}_{\rm{Au+Au}}},
\end{align}
at the same centrality~\cite{Giacalone:2021uhj}.
Figure~\ref{fig:c34ratio}(b) shows the initial-state ratio \(|R_{c_{3,\varepsilon}\{4\}}|\) from \texttt{T\raisebox{-.5ex}{R}ENTo} simulations. In the linear-response limit,
\begin{equation}
R_{c_{3,v}\{4\}} = \frac{\kappa_{3,\rm U}^4}{\kappa_{3,\rm Au}^4} R_{c_{3,\varepsilon}\{4\}} .
\end{equation}
This ratio exhibits the same dependence on $\langle \beta_\mathrm{3,U}^4 \rangle$ as the absolute values $|c_{3,\varepsilon}\{4\}|$ shown in Fig.~\ref{fig:c34ratio}(a). For the ultra-central $^{238}$U+$^{238}$U and $^{197}$Au+$^{197}$Au collisions considered here, the response coefficients are expected to be similar, so that the ratio is mainly controlled by the initial-state eccentricity cumulants~\cite{Noronha-Hostler:2015dbi}.

\begin{figure}[t]
\centering
\includegraphics[width=0.75\linewidth]{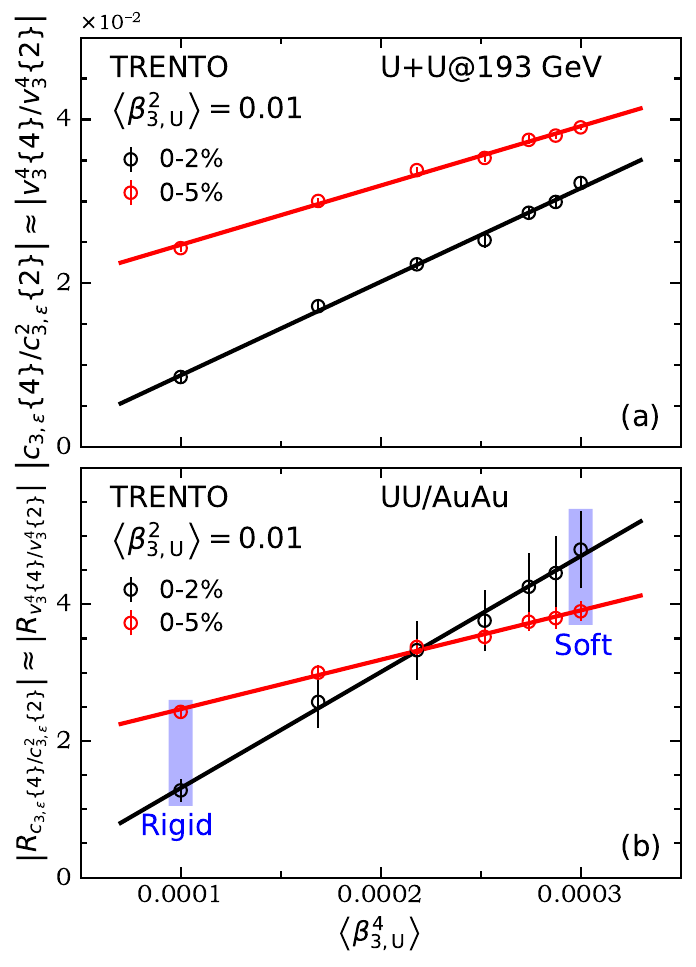}
\vspace{-0.2cm}
\caption{(a) Dependence of the normalized cumulant ratio $|c_{3,\varepsilon}\{4\}/c_{3,\varepsilon}^2\{2\}|$ on the fourth-order moment $\langle \beta_\mathrm{3,U}^4 \rangle$ from \texttt{T\raisebox{-.5ex}{R}ENTo} model in two different centrality intervals, 0--2\% (black symbols) and 0--5\% (red symbols). (b) The ratio between $^{238}$U+$^{238}$U and $^{197}$Au+$^{197}$Au collisions, $|R_{c_{3,\varepsilon}\{4\}/c_{3,\varepsilon}^2}|$, as a function of $\langle \beta_\mathrm{3,U}^4 \rangle$. Linear fits are shown as solid lines. The light blue bands depict exclusively the rigid and soft deformation cases.}
\label{fig:c34overc32ratio}
\end{figure}

The linear-response assumption gives
\begin{align}
v_n\{2\} &= \kappa_n \varepsilon_n\{2\},\\
v_n\{4\} &= \kappa_n \varepsilon_n\{4\},
\label{eq:linearresponse}
\end{align}
and thus \(c_{n,v}\{4\}=\kappa_n^4 c_{n,\varepsilon}\{4\}\). 
For triangular flow in central collisions, this approximation is expected to be reliable because \(v_3\) is mainly generated by the initial triangularity and nonlinear mode-coupling effects are small~\cite{Bhalerao:2014xra,STAR:2022gki}. 
Nevertheless, we treat this mapping as an explicit assumption of the present initial-state study.

A useful observable is the normalized cumulant, for which the leading response coefficient cancels:
\begin{equation}
-\mathrm{nc}_{3,\varepsilon}\{4\}
\equiv
-\frac{c_{3,\varepsilon}\{4\}}{c_{3,\varepsilon}^2\{2\}}
\approx
-\frac{c_{3,v}\{4\}}{c_{3,v}^2\{2\}}
=
\frac{v_3^4\{4\}}{v_3^4\{2\}} .
\end{equation}
Figure~\ref{fig:c34overc32ratio}(a) shows that 
\(|c_{3,\varepsilon}\{4\}/c_{3,\varepsilon}^2\{2\}|\) also increases approximately linearly with 
\(\langle \beta_{\mathrm{3,U}}^4 \rangle\). 
The corresponding system ratio 
\(|R_{c_{3,\varepsilon}\{4\}/c_{3,\varepsilon}^2\{2\}}|\), shown in Fig.~\ref{fig:c34overc32ratio}(b), exhibits a similar linear dependence. 
We have verified that \(^{96}\mathrm{Zr}+^{96}\mathrm{Zr}\) and \(^{96}\mathrm{Ru}+^{96}\mathrm{Ru}\) collisions yield qualitatively identical results.

The previous result is very intuitive. We argue that this phenomenon is a robust approach for extracting both the second- and fourth-order moments of $\beta_{\rm 3,U}$, namely $\langle \beta_\mathrm{3,U}^2 \rangle$ and $\langle \beta_\mathrm{3,U}^4 \rangle$, through measurements of $v_3\{2\}$ and the normalized ratio $v_3\{4\}/v_3\{2\}$, respectively. It is anticipated that this approach will remain valid in full final-state hydrodynamic calculations, and the $R_{\mathcal{O}}$ also largely cancel out final-state effects~\cite{Zhang:2022fou}. The former has already been constrained in a previous RHIC-STAR measurement~\cite{STAR:2025vbp,Zhang:2025hvi}. Once $\langle \beta_\mathrm{3,U}^2 \rangle$ and $\langle \beta_\mathrm{3,U}^4 \rangle$ are determined, the mean deformation $\bar{\beta}_3$ and its variance $\sigma_{\beta_3}^2$ can be derived via Eqs.~\eqref{eq:beta3var} under assumption of Gaussian fluctuations. 

We note that a full quantitative validation would require event-by-event viscous hydrodynamic simulations. This is computationally demanding for the present observable, since the four-particle cumulant in ultra-central collisions is small and requires very high statistics. In our \texttt{T\raisebox{-.5ex}{R}ENTo} study, we use \(2\times10^8\) minimum-bias events for each deformation setting. We therefore leave the hydrodynamic validation to future work and regard the present results as an initial-state sensitivity study.

For fluctuations beyond the Gaussian approximation, higher-order cumulants of $\beta_{n}$, such as skewness and kurtosis, would be required. To illustrate the role of the fluctuation assumption, we write the fourth-order moment of the octupole deformation in terms of the central moments of the \(\beta_3\) distribution:
\begin{equation}
\left\langle \beta_3^4 \right\rangle
=
\bar{\beta}_3^4
+
6\bar{\beta}_3^2\sigma_{\beta_3}^2
+
4\bar{\beta}_3 \mu_{3,\beta_3}
+
\mu_{4,\beta_3},
\label{eq:beta3fourth_nongaussian}
\end{equation}
where
\(\bar{\beta}_3=\langle\beta_3\rangle\),
\(\sigma_{\beta_3}^2=\langle(\beta_3-\bar{\beta}_3)^2\rangle\), and
\(\mu_{n,\beta_3}=\langle(\beta_3-\bar{\beta}_3)^n\rangle\). 
The standardized skewness and excess kurtosis are
\begin{equation}
\gamma_{1,\beta_3}
=
\frac{\mu_{3,\beta_3}}{\sigma_{\beta_3}^3},
\qquad
\gamma_{2,\beta_3}
=
\frac{\mu_{4,\beta_3}}{\sigma_{\beta_3}^4}-3 .
\end{equation}
Therefore, \(\langle \beta_3^4\rangle\) can depend not only on the mean and variance, but also on possible non-Gaussian features of the deformation distribution. In the Gaussian approximation used in this work, 
\(\gamma_{1,\beta_3}=\gamma_{2,\beta_3}=0\), or equivalently
\(\mu_{3,\beta_3}=0\) and \(\mu_{4,\beta_3}=3\sigma_{\beta_3}^4\). This gives
\begin{equation}
\left\langle \beta_3^4 \right\rangle_{\rm Gauss}
=
\bar{\beta}_3^4
+
6\bar{\beta}_3^2\sigma_{\beta_3}^2
+
3\sigma_{\beta_3}^4 .
\label{eq:beta3fourth_gaussian}
\end{equation}

Higher-order flow cumulants may provide additional sensitivity to these non-Gaussian deformation moments. In the following, we briefly discuss the implications of this phenomenon for higher-order cumulant of anisotropic flow. This would further enhance our understanding of the nuclear structure and fluctuations at the subatomic level~\cite{Borghini:2001vi,Bilandzic:2013kga,Bilandzic:2010jr,Jia:2017hbm}. The sixth- and eighth-order cumulants, $c_{n,\varepsilon}\{6\}$ and $c_{n,\varepsilon}\{8\}$, including the contributions from deformation, can be expressed as
\begin{align}
\label{eq:cn6} 
c_{n,\varepsilon}\{6\} &= \langle \varepsilon_n^6 \rangle - 9 \langle \varepsilon_n^4 \rangle \langle \varepsilon_n^2 \rangle + 12 \langle \varepsilon_n^2 \rangle^3 \nonumber\\
& \approx \langle \varepsilon_{n,0}^6 \rangle - 9 \langle \varepsilon_{n,0}^4 \rangle \langle \varepsilon_{n,0}^2 \rangle + 12 \langle \varepsilon_{n,0}^2 \rangle^3 \nonumber\\
&\quad + \left\langle p_n^3 p_n^{3*} \right\rangle \langle \beta_{n}^6 \rangle - 9 \left\langle p_n^2 p_n^{2*} \right\rangle  \left\langle p_n p_n^{*} \right\rangle\langle \beta_{n}^4 \rangle \langle \beta_{n}^2 \rangle \nonumber\\
&\quad+ 12  \left\langle p_n p_n^{*} \right\rangle^3\langle \beta_{n}^2 \rangle^3,\\
\label{eq:cn8} 
c_{n,\varepsilon}\{8\} &= \langle \varepsilon_n^8 \rangle - 16 \langle \varepsilon_n^6 \rangle \langle \varepsilon_n^2 \rangle - 18 \langle \varepsilon_n^4 \rangle^2 \nonumber\\
&\quad + 144 \langle \varepsilon_n^4 \rangle \langle \varepsilon_n^2 \rangle^2 - 144 \langle \varepsilon_n^2 \rangle^4\nonumber\\
& \approx \langle \varepsilon_{n,0}^8 \rangle - 16 \langle \varepsilon_{n,0}^6 \rangle \langle \varepsilon_{n,0}^2 \rangle - 18 \langle \varepsilon_{n,0}^4 \rangle^2 \nonumber\\
&\quad + 144 \langle \varepsilon_{n,0}^4 \rangle \langle \varepsilon_{n,0}^2 \rangle^2 - 144 \langle \varepsilon_{n,0}^2 \rangle^4\nonumber\\
& \quad+\left\langle p_n^4 p_n^{4*} \right\rangle\langle \beta_{n}^8 \rangle - 16 \left\langle p_n^3 p_n^{3*} \right\rangle\left\langle p_n p_n^{*} \right\rangle\langle \beta_{n}^6 \rangle \langle \beta_{n}^2 \rangle \nonumber\\
&\quad - 18 \left\langle p_n^2 p_n^{2*} \right\rangle^2\langle \beta_{n}^4 \rangle^2 \nonumber\\
&\quad + 144 \left\langle p_n^2 p_n^{2*} \right\rangle\left\langle p_n p_n^{*} \right\rangle^2 \langle \beta_{n}^4 \rangle \langle \beta_{n}^2 \rangle^2 \nonumber\\
&\quad - 144 \left\langle p_n p_n^{*} \right\rangle^4 \langle \beta_{n}^2 \rangle^4.
\end{align}

We observe that $c_{n,\varepsilon}\{6\}$ in Eq.~\eqref{eq:cn6} depends linearly on the sixth-order moment $\langle \beta_n^6 \rangle$ when the second- and fourth-order moments, $\langle \beta_n^2 \rangle$ and $\langle \beta_n^4 \rangle$, are held fixed. Similarly, $c_{n,\varepsilon}\{8\}$ in Eq.~\eqref{eq:cn8} exhibits linear dependence on the eighth-order moment $\langle \beta_n^8 \rangle$ when the second-, fourth-, and sixth-order moments $\langle \beta_n^2 \rangle$, $\langle \beta_n^4 \rangle$, and $\langle \beta_n^6 \rangle$ are fixed. 

However, we emphasize that these relations do not by themselves constitute a closed, model-independent extraction of all non-Gaussian parameters of the deformation distribution. Once the Gaussian approximation is relaxed, additional unknowns, such as skewness, kurtosis, and higher central moments of $\beta_n$, enter the problem. Moreover, higher-order flow cumulants may also receive contributions from higher-order response coefficients and nonlinear response terms. Therefore, extracting the full non-Gaussian deformation distribution from two-, four-, six-, and eight-particle cumulants requires additional assumptions, such as a controlled truncation of the response expansion, or independent constraints from nuclear-structure calculations and realistic event-by-event hydrodynamic simulations.

In the present work, the main extraction strategy is based on the Gaussian approximation, under which the deformation distribution is fully characterized by $\bar{\beta}_3$ and $\sigma_{\beta_3}^2$. The higher-order cumulant expressions in Eqs.~\eqref{eq:cn6} and \eqref{eq:cn8} are provided to illustrate how future measurements could, in principle, test deviations from Gaussianity and constrain higher moments of the deformation distribution, provided that the higher-order response contributions are either negligible or independently constrained.

\textit{Summary and outlook.---} In summary, we have developed a scaling approach to probe nuclear shape fluctuations using multi-particle flow cumulants in relativistic heavy-ion collisions, and have applied it to study octupole deformation and its fluctuations in $^{238}$U. By deriving analytical expressions for $c_{3,\varepsilon}\{2\}$ and $c_{3,\varepsilon}\{4\}$ and validating them with the \texttt{T\raisebox{-.5ex}{R}ENTo} model, we demonstrated that, i) The two-particle cumulant $c_{3,\varepsilon}\{2\}$ is sensitive only to the second-order moment $\left\langle\beta_3^2\right\rangle$ of the deformation; ii) the four-particle cumulant ratio $|R_{c_{3,\varepsilon}\{4\}}|$ and its normalized ratio $|R_{c_{3,\varepsilon}\{4\}/c_{3,\varepsilon}^2\{2\}}|$ are approximately linearly scale to the fourth-order moment $\langle \beta_3^4 \rangle$ at fixed $\langle \beta_3^2 \rangle$. Under the assumption of Gaussian fluctuations, by mean of Eqs.~\eqref{eq:beta3var}, this approach enables the simultaneous extraction of both the mean octupole deformation $\bar{\beta}_3$ and its variance $\sigma^2_{\beta_3}$ from experimentally accessible flow cumulants.

We demonstrate that the low-energy octupole collective mode directly underlies the high-energy triangular flow fluctuations. It offers a means to experimentally distinguish between soft (vibrational) and rigid (static) deformation mechanisms in nuclei. We also briefly outline the implications for future studies of non-Gaussian fluctuations via higher-order cumulants. Our study paves a novel pathway to further understand nuclear shape fluctuations, constrain the initial conditions of QGP droplets. This work provides a concrete path to determine ground-state nuclear shapes in experiments at RHIC and the LHC.

\textit{Acknowledgements.---} We thank Weiyao Ke for valuable discussions. This work is supported in part by the National Key Research and Development Program of China under Contract Nos. 2024YFA1612600 and 2022YFA1604900, the National Natural Science Foundation of China (NSFC) under Contract Nos. 12025501, 12547102, 12205051, the Natural Science Foundation of Shanghai under Contract No. 23JC1400200, Shanghai Pujiang Talents Program under Contract No. 24PJA009, China Postdoctoral Science Foundation under Grant No. 2024M750489. J. Jia is supported by the U.S. Department of Energy, Office of Science, Office of Nuclear Physics, under DOE Awards No. DE-SC0024602.

\bibliography{ref}
\end{document}